\documentclass[conference]{IEEEtran}

\IEEEoverridecommandlockouts

\usepackage{cite}
\usepackage{algorithmic}
\usepackage{graphicx}
\usepackage{textcomp}
\usepackage{xcolor}

\usepackage[cmex10]{amsmath}
\usepackage{amssymb}
\usepackage{amsthm}
\usepackage{amsfonts}
\usepackage[colorlinks=false, pdfborder={0 0 0}]{hyperref}
\usepackage[capitalise]{cleveref}

\usepackage{url}
\usepackage{enumerate}
\usepackage[inline]{enumitem}
\usepackage{booktabs}
\usepackage{multirow}
\usepackage{colortbl}
\usepackage[font=footnotesize]{caption}
\usepackage[font=footnotesize]{subcaption}

\usepackage[T1]{fontenc}

\usepackage{tikz}
\usepackage{tikz-qtree}
\usepackage{tikz-3dplot}
\tdplotsetmaincoords{70}{20}
\usetikzlibrary{backgrounds,fit}
\usetikzlibrary{matrix,calc,arrows}

\newtheorem{lemma}{Lemma}
\newtheorem{theorem}{Theorem}

\theoremstyle{definition}\newtheorem{definition}{Definition}
\theoremstyle{remark}\newtheorem{example}{Example}
\theoremstyle{remark}

\newcommand{\ellmax}{\ell^\text{max}}
\newcommand{\ie}{\text{i.e.}}

\begin{document}

\title{On Multi-Channel Huffman Codes for Asymmetric-Alphabet Channels}

\author{
	\IEEEauthorblockN{Hoover~H.~F.~Yin, Xishi~Wang, Ka~Hei~Ng, Russell~W.~F.~Lai, Lucien~K.~L.~Ng, and Jack~P.~K.~Ma}
	\thanks{
        H.~Yin is with the Institute of Network Coding, The Chinese University of Hong Kong.
		X.~Wang, L.~Ng and J.~Ma are with the Department of Information Engineering, The Chinese University of Hong Kong.
		K.~Ng is with the Department of Physics, The Chinese University of Hong Kong.
		R.~Lai is with the Chair of Applied Cryptography, Friedrich-Alexander-Universit\"at Erlangen-N\"urnberg. He is supported by the State of Bavaria at the Nuremberg Campus of Technology (NCT).
	}
}

\maketitle

\begin{abstract}
Zero-error single-channel source coding has been studied extensively over the past decades.
Its natural multi-channel generalization is however not well investigated.
While the special case with multiple symmetric-alphabet channels was studied a decade ago,
codes in such setting have no advantage over single-channel codes in data compression, making them worthless in most applications.
With essentially no development since the last decade, in this paper, we break the stalemate by showing that it is possible to beat single-channel source codes in terms of compression assuming asymmetric-alphabet channels.
We present the multi-channel analog of several classical results in single-channel source coding, such as that a multi-channel Huffman code is an optimal tree-decodable code.
We also show some evidences that finding an efficient construction of multi-channel Huffman codes may be hard.
Nevertheless, we propose a suboptimal code construction whose redundancy is guaranteed to be no larger than that of an optimal single-channel source code.
\end{abstract}

\section{Introduction}

Zero-error source coding is one of the oldest branches in information theory.
In a traditional (single-channel) source coding problem, a transmitter wishes to send an information source to a receiver with zero error through an error-free channel.
To better utilize the channel, it is desirable to encode the information source in such a way that minimizes the number of symbols transmitted through the channel while ensuring correct decoding.
In technical terms, the goal is to minimize the expected codeword length of a source code for an information source.
Given an arbitrary information source, the well-known Huffman procedure \cite{huffman} can produce an optimal\footnote{%
	In this paper, the optimality we considered is for symbol-by-symbol coding with known probability masses of an information source.
}
code known as a Huffman code.
On the other hand, the entropy bound shows that the information theoretical lower bound of the expected codeword length of a source code is the entropy of the source information.
Despite the optimality of the Huffman code,
it does not achieve the entropy bound in general.

As a natural generalization, zero-error multi-channel source coding was proposed in \cite{yao10} where each source symbol is mapped to a codeword which spreads across multiple channels.
This problem is more complicated than its single-channel counterpart,
as the sequence of received symbols in individual channels are not necessary uniquely decodable, even though the overall source code is uniquely decodable.
In \cite{yao10}, only the case where all channels use the same alphabet size, \ie, \emph{symmetric-alphabet channels}, was investigated.
Among many other results, \cite{yao10} showed that
a multi-channel code (for symmetric-alphabet channels) can be equated to a single-channel code, so that the existing tools of source coding can be applied to analyze the multi-channel code.
However, this implies that such a multi-channel code cannot improve the compression ability.
It is thus natural to ask: What if at least one channel has an alphabet size different from that of another channel, \ie, we have \emph{asymmetric-alphabet channels}?
Are there any new opportunities and challenges under this extended setting?

\subsection{Motivating Examples}

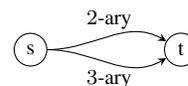
\begin{figure}
	\footnotesize
	\centering
	\begin{tikzpicture}[scale=.7]
		\node[draw,circle] (s) {s};
		\node[draw,circle,node distance=2cm,right of=s] (t) {t};
		\draw[-stealth] (s.east) to [out=0,in=150] node[above] {$2$-ary} (t);
		\draw[-stealth] (s.east) to [out=0,in=210] node[below] {$3$-ary} (t);
	\end{tikzpicture}
	\vskip -.7em
	\caption{A $2$-channel zero-error communication system where the two channels use a binary and a ternary alphabets respectively.}
	\label{fig:network}
\end{figure}

Our starting point is the observation that, in some cases, a multi-channel code with asymmetric alphabets achieves a better compression ability than an optimal single-channel code.
Specifically, we consider the network in \cref{fig:network} and compare the expected codeword (description) length\footnote{The description length of a codeword is the number of information unit (e.g., nat) required to represent this codeword.} of an optimal $(2, 3)$-ary tree-decodable code\footnote{A multi-channel prefix-free code might not have a decoding tree. We defer the detailed discussion regarding this issue to \cref{sec:tree}.}, the binary Huffman code, and the ternary Huffman code for the information sources with probability masses $\{1/6, 1/6, 1/3, 1/3\}$ and $\{1/6, 1/6, 1/6, 1/2\}$ respectively in the following examples.
For convenience, we use the information unit nat (base $e$, the Euler's number) throughout this paper.

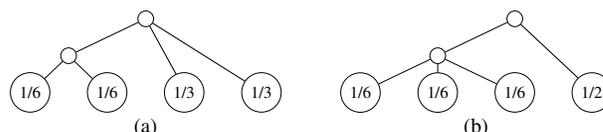
\begin{figure}
    \small
    \centering
    \begin{subfigure}{.23\textwidth}
        \centering
        \begin{tikzpicture}[scale=.7,every tree node/.style={draw,circle},level distance=.7cm,sibling distance=.7cm,edge from parent path={(\tikzparentnode)--(\tikzchildnode)}]
        \Tree [.\node(r){};
        [.{}
        [.{1/6} ] [.{1/6} ]
        ]
        \edge[draw=none];
        [
        \edge[draw=none];
        [.\node(a){1/3}; ]
        \edge[draw=none];
        [.\node(b){1/3}; ]
        ]
        ];
        \draw (r) -- (a);
        \draw (r) -- (b);
        \end{tikzpicture}
		\vskip -.5em
        \caption{}
        \label{fig:optimal_eg1}
    \end{subfigure}%
    ~
    \begin{subfigure}{.23\textwidth}
        \centering
        \begin{tikzpicture}[scale=.7,every tree node/.style={draw,circle},level distance=.7cm,sibling distance=.7cm,edge from parent path={(\tikzparentnode)--(\tikzchildnode)}]
        \Tree [.\node(r){};
        [.{}
        [.{1/6} ] [.{1/6} ] [.{1/6} ]
        ]
        \edge[draw=none];
        [
        \edge[draw=none];
        [.\node(a){1/2}; ]
        ]
        ];
        \draw (r) -- (a);
        \end{tikzpicture}
		\vskip -.5em
        \caption{}
        \label{fig:optimal_eg2}
    \end{subfigure}
	\vskip -.7em
    \caption{
        Examples of optimal $(2, 3)$-ary codes achieving the entropy bound.
    }
\end{figure}

\begin{example}
    \label{ex:better_compression1}
    In \cref{fig:optimal_eg1}, the codeword lengths for the $1/6$ and $1/3$ masses are $(\ln 2 + \ln 3)$ nats and $\ln 3$ nats respectively.
    The expected codeword length of the $(2, 3)$-ary code is $(\frac13 \ln 2 + \ln 3)$ nats, which is the entropy of the information source.
    The expected codeword length of the corresponding binary and ternary Huffman codes are $2 \ln 2$ nats and $\frac43 \ln 3$ nats respectively.
\end{example}

\begin{example}
    \label{ex:better_compression2}
    In \cref{fig:optimal_eg2}, the codeword lengths for the $1/6$ and $1/2$ masses are $(\ln 2 + \ln 3)$ nats and $\ln 2$ nats respectively.
    The expected codeword length is $(\ln 2 + \frac12 \ln 3)$ nats, which is the entropy of the information source.
    The expected codeword length of the corresponding binary and ternary Huffman codes are $\frac{11}{6} \ln 2$ nats and $\frac43 \ln 3$ nats respectively.
\end{example}

\begin{table}
    \centering
    \caption{Expected codeword lengths of optimal codes in nats}
    \label{tab:entropy}
	\vskip -.6em
    \begin{tabular}{r|c|c|c}
        & $(2, 3)$-ary & $2$-ary & $3$-ary \\ \midrule
        \cref{fig:optimal_eg1} & 1.32966134885 & 1.38629436112 & 1.46481638489\\
        \cref{fig:optimal_eg2} & 1.24245332489 & 1.27076983103 & 1.46481638489
    \end{tabular}
\end{table}

We summarize the above two examples and give the numerical values of the expected codeword length in \cref{tab:entropy}.
We can see that the expected codeword length when both channels are in used is shorter than the other two single-channel cases,
\ie, the optimal multi-channel source code outperforms the optimal single-channel source codes.

\subsection{Our Contributions}

Given the above examples, it is natural to ask if a multi-channel generalization of the Huffman procedure systematically produces an optimal code.
We give a partial affirmative answer to the above question.
In particular, we formulate the generalized Huffman procedure and show that it indeed produces an optimal tree-decodable code.
The generalized Huffman procedure follows the same idea of iteratively merging the smallest probability masses into bigger ones, until reaching probability $1$.
However, unlike the single-channel case, there is more than one choice of the number of masses to be merged in each iteration.
This results in an exponential number of potential merge sequences, and only choosing a ``correct'' merging step in every iteration results in an optimal code.
It is therefore unclear how the generalized Huffman procedure can be computed in polynomial time.

Towards designing an algorithm which computes or approximates the generalized Huffman procedure in polynomial time, we devise and investigate several heuristics for pruning the merging steps.
We first investigate pruning by some natural measures such as (our multi-channel generalization of) local redundancy, expected codeword length, and entropy.
Unfortunately none of these simplistic strategies can guarantee the production of an optimal code.

One of our results is the following pruning strategy based on iterative executions of the (single-channel) Huffman procedure:
For a multiset of probability masses either given initially or produced by the previous iteration, we run the Huffman procedure on these masses with respect to the alphabet size of each channel.
Let $q$ be the alphabet size of the resulting Huffman code with the smallest expected length.
We merge the $q$ smallest probability masses into one and proceed to the next round.

Since all single-channel Huffman codes are included as potential choices of the above minimizing procedure, the resulting code guarantees a compression ability no worse than any of the optimal single-channel codes.
While we are unable to show that this heuristic produces an optimal tree-decodable code, we have yet to find a counterexample either.
We leave it as an open problem to prove the (sub)optimality of the resulting code.
We remark that even if the heuristic produces an optimal tree-decodable code, it is unclear if the code is also an optimal prefix code except for the $2$-channel case.
It is also unclear if an optimal prefix code is an optimal uniquely decodable code.

\subsection{Paper Organization}

We first introduce the background of multi-channel source coding in \cref{sec:bg}.
Then, we present some generalizations of some well-known results for single-channel source coding in \cref{sec:general} and the multi-channel Huffman procedure for asymmetric-alphabet channels in \cref{sec:huffman}.
We show some evidence that an efficient construction of multi-channel Huffman codes is not trivial and propose a suboptimal code construction in \cref{sec:approx}.
Lastly, we conclude the paper and propose some future directions in \cref{sec:conclude}.

\section{Background}
\label{sec:bg}

Unless otherwise specified, we use the notation defined in this section throughout this paper.
For $n \in \mathbb{Z}^+$, define $[n] = \{1, 2, \ldots, n\}$.

\subsection{Multi-Channel Source Coding}

The definition of a multi-channel source code was proposed in \cite{yao10} for symmetric-alphabet channels and extended in \cite{packing} for asymmetric-alphabet channels.
Consider an $n$-channel system where the $i$th channel uses an alphabet $\mathcal{Z}_i$ of size $q_i$, $i \in [n]$.
Without loss of generality, we assume $q_1 \leq q_2 \leq \ldots \leq q_n$ in this paper.
Let $\mathcal{Z}$ be the alphabet of the information source $Z$.
Denote by $\epsilon$ the empty string.
Define $\mathcal{Z}_i^0 = \{\epsilon\}$ and $\mathcal{Z}_i^j = \{wv \colon w \in \mathcal{Z}_i^{j-1}, v \in \mathcal{Z}_i\}$ for $j \in \mathbb{Z}^+$.
That is, $\mathcal{Z}_i^j$ contains all the strings of length $j$ over the alphabet $\mathcal{Z}_i$.
The set of all the strings of countable length over the alphabet $\mathcal{Z}_i$ is denoted by $\mathcal{Z}_i^\ast = \bigcup_{j = 0}^\infty \mathcal{Z}_i^j$.
Every element in $\prod_{i = 1}^n \mathcal{Z}_i^\ast$ is called a \emph{word}.

\begin{definition}[Multi-Channel Source Codes]
    A $(q_1,\ldots,q_n)$-ary \emph{source code} $\mathcal{Q}$ for the information source $Z$ is a mapping from $\mathcal{Z}$ to $\prod_{i = 1}^n \mathcal{Z}_i^\ast$.
\end{definition}

For any \emph{source symbol} $z \in \mathcal{Z}$, the ordered $n$-tuple $\mathcal{Q}(z)$ is the \emph{codeword} for $z$.
The $i$th component of the codeword is transmitted through the $i$th channel.
When we transmit more than one codeword, the codewords are concatenated channel-wise, \ie, the boundaries of the codewords are not explicit.

\begin{definition}[Uniquely Decodable Codes]
	If the finite sequences of codewords of any two distinct finite sequences of source symbols are different, then the source code is called a \emph{uniquely decodable code}.
\end{definition}

\begin{definition}[Multi-Channel Prefix-Free Codes]
    \label{def:prefix}
	Two words are \emph{prefix free} (to each other) if there exists at least one channel $i$ such that the $i$th component of the two words are prefix free.
    A $(q_1,\ldots,q_n)$-ary \emph{prefix-free code} is a $(q_1,\ldots,q_n)$-ary source code such that every pair of codewords are prefix free.
\end{definition}

By convention, a prefix-free code is also called a \emph{prefix code}.
Any codeword of a uniquely decodable multi-channel source code can be decoded without referring to the symbols of any future codewords if and only if the code is a multi-channel prefix code \cite{yao10,packing}.

\subsection{Single-Channel Representation}

As a channel does not interfere the symbols in another channel, we can express a multi-channel word as a single-channel word with a dynamic alphabet.
For example,
let $\mathcal{Z}_1 = \{0, 1\}$ and $\mathcal{Z}_2 = \{a, b, c\}$.
A $2$-channel word $(01, bca)$ can be expressed as either $01bca$, $0b1ca$, $0bc1a$, $0bca1$, $b01ca$, $b0c1a$, $b0ca1$, $bc01a$, $bc0a1$, or $bca01$.
Note that $\mathcal{Z}_1$ and $\mathcal{Z}_2$ have different letter costs.

However, this single-channel representation is different from a single-channel word with unequal letter cost, a concept proposed in \cite{unequal}.
For example,
in a single-channel code with unequal letter cost for the alphabet $\{0, 1, a, b, c\}$, the words $0$ and $a$ are prefix-free.
However, the $2$-channel words $(0,\epsilon)$ and $(\epsilon,a)$ are not prefix-free although their single-channel representations are $0$ and $a$ respectively.

When the alphabet sizes of the channels are the same, we can see that the single-channel representation of a multi-channel code can be reduced to a single-channel code with equal letter cost.
That is why the compression ability cannot be enhanced in the setting proposed in \cite{yao10}.

\subsection{Kraft Inequality and Entropy Bound}

Let $n$ and $m$ be the number of channels and number of codewords respectively.
Denote by $H(\cdot)$ the entropy measured in nats.
All results can be trivially generalized to use other bases of logarithm.

The length of the $j$th codeword in the $i$th channel is denoted by $\ell_i^j$.
The length tuple of the $j$th codeword is $(\ell_1^j, \ldots, \ell_n^j)$.
The (description) length of the codeword, defined by $\ell^j := \sum_i \ell_i^j \ln q_i$, is the number of nats required to represent it.
Define $\ellmax_i := \max_{j = 1}^m \ell_i^j$.

The multi-channel analog to Kraft inequality and entropy bound were proposed in \cite{yao10} for symmetric-alphabet channels.
They were extended for asymmetric-alphabet channels in the conference paper \cite{packing} but their proofs were separated in the preprint \cite{packing_arxiv}.
For the sake of completeness, we reproduce the proofs in this paper.

\begin{theorem}[Kraft Inequality \cite{packing}] \label{thm:kraft}
    If $\mathcal{Q}$ is uniquely decodable, then the length tuples of its codewords satisfy
    \begin{equation} \label{eq:kraft}
    \sum_{j = 1}^m \prod_{i = 1}^n q_i^{-\ell_i^j} \le 1.
    \end{equation}
\end{theorem}

\begin{IEEEproof}
    We use a similar technique as in \cite{mcmillan_proof}.
    Let $N$ be an arbitrary positive integer.
    Consider
    \begin{IEEEeqnarray*}{Cl}
        & \left( \sum_{j = 1}^m \prod_{i = 1}^n q_i^{-\ell_i^j} \right)^N = \sum_{j_1 = 1}^m \sum_{j_2 = 1}^m \ldots \sum_{j_N = 1}^m \left( \prod_{i = 1}^n q_i^{-\sum_{k = 1}^N \ell_i^{j_k}} \right) \\
        = & \sum_{k_1 = 1}^{N\ellmax_1} \sum_{k_2 = 1}^{N\ell_2^\text{max}} \ldots \sum_{k_n = 1}^{N\ellmax_n} A_{k_1, k_2, \ldots, k_n} \prod_{i = 1}^n q_i^{-k_i},
        \yesnumber
        \label{eq:kraft_expand}
    \end{IEEEeqnarray*}
    where $A_{k_1, k_2, \ldots, k_n}$ is the coefficient of $\prod_{i = 1}^n q_i^{-k_i}$ in $( \sum_{j = 1}^m \prod_{i = 1}^n q_i^{-\ell_i^j} )^N$.

    Note that $A_{k_1, k_2, \ldots, k_n}$ gives the total number of sequences of $N$ codewords with a total length of $k_i$ symbols in the $i$th channel for all $i = 1, 2, \ldots, n$.
    Since the code is uniquely decodable, these code sequences must be distinct.
    So, the number $A_{k_1, k_2, \ldots, k_n}$ must be no more than the total number of distinct sequences where there are $k_i$ symbols in the $i$th channel for all $i = 1, 2, \ldots, n$.
    That is, we have
    $A_{k_1, k_2, \ldots, k_n} \le \prod_{i = 1}^n q_i^{k_i}$.
    By substituting this inequality into \eqref{eq:kraft_expand}, we get
    \begin{equation*}
    \sum_{j = 1}^m \prod_{i = 1}^n q_i^{-\ell_i^j} \le \left(\sum_{k_1 = 1}^{N\ellmax_1} \sum_{k_2 = 1}^{N\ell_2^\text{max}} \ldots \sum_{k_n = 1}^{N\ellmax_n} 1\right)^{1/N}.
    \end{equation*}
    Since this inequality holds for any $N$, we let $N \to \infty$ and obtain our desired Kraft inequality.
\end{IEEEproof}

\begin{theorem}[Entropy Bound \cite{packing}] \label{thm:entropy}
    Let $\mathcal{Q}$ be a uniquely decodable code for a source random variable $Z$ with probability $\{p_1, p_2, \ldots, p_m\}$ and entropy $H(Z)$.
    Then,
    \begin{equation} \label{eq:entropy}
    \sum_{j = 1}^m p_j \sum_{i = 1}^n \ell_i^j \ln q_i \ge H(Z),
    \end{equation}
    where the equality holds if and only if $\sum_{i = 1}^n \ell_i^j \ln q_i = -\ln p_j$.
\end{theorem}

\begin{IEEEproof}
    Here we use a standard technique.
    Consider
    \begin{IEEEeqnarray*}{Cl}
        & \sum_{j = 1}^m p_j \sum_{i = 1}^n \ln q_i^{\ell_i^j} - H(Z)\\
        = & \sum_{j = 1}^m p_j \ln \left( p_j \prod_{i = 1}^n q_i^{\ell_i^j} \right)\\
        \ge & \sum_{j = 1}^m p_j \left( 1 - \left( p_j \prod_{i = 1}^n q_i^{\ell_i^j} \right)^{-1} \right) \yesnumber \label{eq:entropy_ln} \\
        = & \left( 1 - \sum_{j = 1}^m \prod_{i = 1}^n q_i^{-\ell_i^j} \right) \ge 0, \yesnumber \label{eq:entropy_kraft}
    \end{IEEEeqnarray*}
    where \eqref{eq:entropy_ln} follows the inequality $\ln a \ge 1 - 1/a$ for any $a > 0$, and \eqref{eq:entropy_kraft} follows the multi-channel Kraft inequality.

    The equality in \eqref{eq:entropy_ln} holds if and only if $p_j \prod_{i = 1}^n q_i^{\ell_i^j} = 1$ for all $j$, or equivalently, $\sum_{i = 1}^n \ln q_i^{\ell_i^j} = -\ln p_j$ for all $j$.
    Under this condition, we have
    $\sum_{j = 1}^m \prod_{i = 1}^n q_i^{-\ell_i^j} = \sum_{j = 1}^m p_j = 1$.
    So, the equality in \eqref{eq:entropy_kraft} also holds, which means that the bound is tight.
\end{IEEEproof}

The single-channel Kraft inequality gives a necessary and sufficient condition for the existence of a single-channel uniquely decodable code.
However in the multi-channel analog, the sufficiency does not hold in general.
The reason can be easily seen from the rectangle packing formulation for prefix codes in \cite{packing}: The Kraft inequality only states that the sum of areas of blocks (codewords) is no larger than the area of the container (all possible words). The geometry of the blocks (the packability of the blocks in the container) is not captured.
On the other hand, the interpretation of the entropy bound is consistent with the traditional version that the expected codeword (description) length is no less than the entropy.

\subsection{Multi-Channel Tree-Decodable Codes}
\label{sec:tree}

A single-channel prefix code can be represented by a decoding tree. 
Each branch of an internal node (including the root node) corresponds to a symbol in the alphabet.
When we decode a codeword, we traverse the tree from the root node.
We read the symbols in the codeword one by one.
For each symbol, we traverse through the corresponding branch.
We can decode a codeword once we reach a leaf of the tree.

When it comes to the multi-channel case, each internal node is assigned to a channel.
An internal node belongs to \emph{class} $i$ if it is assigned to the $i$th channel, and is only allowed to have at most $q_i$ children.
The generalized depth of a leaf is defined as the length of the codeword that the leaf corresponds to.
See \cref{fig:optimal_eg1,fig:optimal_eg2} as examples of a multi-channel decoding tree.
A codeword can be decoded by traversing the tree.
Suppose the current node being traversed is of class $i$, we read a symbol from the $i$th channel of the codeword and traverse to the corresponding branch.
Clearly decoding is sequential, \ie, without referring to the symbols of any future codewords.
A code which has a decoding tree is called a \emph{tree-decodable code}.
The decoding tree of an optimal tree-decodable code is called an \emph{optimal decoding tree}.
Note that a tree-decodable code can have more than one decoding tree.

A tree-decodable code is a prefix code \cite{yao10}.
It is because for any two distinct leaves in a decoding tree, the codewords are not prefix of each other on the channel their lowest common ancestor corresponds to.
The converse is not true in general for $n > 2$ channels.
For each channel, if there is a codeword of a prefix code having an empty string in this channel, then we cannot draw a decoding tree for the code.
Below is an example given in \cite{yao10}.
\begin{example} \label{eg:3d}
    Consider a binary $3$-channel source code, \ie, $q_1 = q_2 = q_3 = 2$.
    The codewords $\{(0,0,\epsilon), (1,\epsilon,0), (\epsilon,1,1)\}$ form a prefix code which is not tree-decodable.
\end{example}

The $2$-channel setting is special in the sense that there cannot exist a codeword having an empty string in the first channel, and another codeword having an empty string in the second channel, since they are not prefix-free. With a more careful argument, we have the following theorem.

\begin{theorem}
    Any $2$-channel prefix code is tree-decodable.
\end{theorem}

\begin{IEEEproof}
    We first argue that, for any $2$-channel prefix code with at least two codewords, there exists at least one channel $i \in \{1,2\}$ such that for all codewords the $i$th component is not an empty string.
    We call such channel $i$ always non-empty.
    Suppose not, then there exist codewords $(c_1, \epsilon)$ and $(\epsilon, c_2)$ for some strings $c_1, c_2$.
    Since $\epsilon$ is a prefix of both $c_1$ and $c_2$, they are not prefix-free.

    Next, given a $2$-channel prefix code, we construct a decoding tree as follows.
    If the code has only one codeword, then we construct a tree with a single node associated to this codeword.
    Otherwise, suppose there are at least 2 codewords.
    By the argument above, there exists an always non-empty channel $i$.
    Let the root node be of class $i$.
    We partition the codewords depending on the first symbol of the codewords in the $i$th channel into the chunks $P_1, \ldots, P_{q_i}$.
    For each chunk $P_j$, we remove the first symbol of the codewords in the $i$th channel.
    This results in a sub-code $\tilde P_j$.

    We argue that $\tilde P_j$ is prefix-free for all $j$.
    Suppose not, let $(c_1,c_2)$ and $(c_1',c_2')$ be codewords in $\tilde P_j$ which are not prefix-free.
    Without loss of generality, suppose channel 1 was used for partitioning.
    Then before the truncation the codewords were $(a \| c_1, c_2)$ and $(a \| c_1', c_2')$ for some symbol $a$.
    Clearly, since $(c_1,c_2)$ and $(c_1',c_2')$ are not prefix-free, $(a \| c_1, c_2)$ and $(a \| c_1', c_2')$ are also not prefix-free.
    This contradicts to the assumption that we start with a prefix code.

    Since each $\tilde P_j$ is prefix-free, we can run the above procedure on $\tilde P_j$ recursively and link the resulting root node to the root node constructed above with the $j$th edge.
    This results in a decoding tree.
\end{IEEEproof}

\section{General Results on Multi-Channel Source Codes}
\label{sec:general}

Towards a better general understanding of multi-channel source codes, we state a collection of elementary results regarding multi-channel source / tree-decodable codes which were not explicitly stated before. These results could help defining, analyzing, and approximating the multi-channel Huffman procedure in the remaining sections. These results are not necessary for understanding the multi-channel Huffman procedure, and can be skipped if the reader desires.

\subsection{Multi-Channel is No Worse than Single-Channel}

\begin{definition} \label{def:ext}
    The \emph{trivial $n$-channel extension} of a single-channel source code expends each codeword into a $n$-tuple where all the new components are empty strings.
\end{definition}

\begin{example}
    If the channel used by a single-channel source code is the $2$nd channel of a $3$-channel system, then the trivial $3$-channel extension of the code is to substitute every codeword $c$ into $(\epsilon, c, \epsilon)$.
\end{example}

We first state the following trivial but important result by viewing $q_i$-ary source codes as special cases of $(q_1,\ldots,q_n)$-ary source codes, which suggests the worthiness of studying multi-channel source codes.

\begin{theorem} \label{thm:trivial}
    The expected codeword length of an optimal $(q_1,\ldots,q_n)$-ary source code is no worse than that of an optimal $q_i$-ary source code for all $i \in [n]$.
\end{theorem}

\begin{IEEEproof}
    The trivial multi-channel extension of an optimal single-channel source code is a multi-channel source code.
    Note that the length of every codeword is conserved, so the expected codeword length of an optimal multi-channel source code is no worse than the one of the above trivial multi-channel extension.
\end{IEEEproof}

\subsection{Source Coding Theorem for Symbol Codes}

Let $L_\text{Huff}$ be the expected codeword length of an optimal $q$-ary single-channel uniquely decodable code.
One of the classical results by Shannon \cite{shannon} which is known as the source coding theorem for symbol codes is that $H(Z) \le L_\text{Huff} < H(Z)+\ln q$.
That is, we can use no more than $\ln q$ nats from the entropy to represent an optimal code.
The following theorem, which is a natural multi-channel analog, extends a similar idea that we can use no more than $\ln q_1$ nats (recall that $q_1 = \min_{i = 1}^n q_i$) from the entropy to represent an optimal code.

\begin{theorem}[Source Coding Theorem for Symbol Codes] \label{thm:source}
	Let $L_\text{opt}$ be the expected codeword length of an optimal $(q_1,\ldots,q_n)$-ary uniquely decodable code.
	Then, $H(Z) \le L_\text{opt} < H(Z) + \ln q_1$.
\end{theorem}

\begin{IEEEproof}
	The first inequality is the entropy bound.
	For the second inequality, we construct an optimal single-channel $q_i$-ary uniquely decodable code for the $i$th channel.
	Let $L_{\text{Huff}_i}$ $q_i$-its be its expected codeword length.
	We have $L_\text{opt} \le L_{\text{Huff}_i} < H(Z) + \ln q_i$, where the first inequality is by \cref{thm:trivial}.
	As we know that $L_\text{opt} \le L_{\text{Huff}_i}$ for all $i$, we have $L_\text{opt} < H(Z) + \ln q_i$ for all $i$, \ie, we have $L_\text{opt} < H(Z) + \ln q_1$.
\end{IEEEproof}

It is also not hard to show that the above upper bound is tight, which is stated in the following theorem.

\begin{theorem} \label{thm:tight}
	$H(Z) + \ln q_1$ is the tightest upper bound on $L_\text{opt}$ which depends only on $H(Z)$.
\end{theorem}

\begin{IEEEproof}
	Consider a multiset of probability masses $P = \{1-(q_1-1)/k\}\uplus(\biguplus_{i = 1}^{q_1-1}\{1/k\})$ where $k \ge q_1$.
	Note that $|P| = q_1$ and $\sum_{x \in P} x = 1$.
	The optimal $n$-channel code is the trivial $n$-channel extension of the optimal $q_1$-ary single-channel code.
	The codeword length is $1$ $q_1$-it, \ie, $\ln q_1$ nats.
	By taking $k \to \infty$, we have $H(Z) \to 0$ so that $L_\text{opt} \to H(Z)+\ln q_1$.
\end{IEEEproof}

\subsection{Local Redundancy of Tree-Decodable Codes}

The concept of local redundancy was introduced in \cite{redundancy}, which is a tool for understanding the redundancy of a tree-decodable code.
An internal node is a non-leaf node in a decoding tree.
In the decoding tree of a single-channel $q$-ary source code, the local redundancy of an internal node $k$ is defined by $r_k = s_k(\ln q -h_k)$ where $s_k$ and $h_k$ are the reaching probability and the entropy of the conditional branching probabilities respectively of node $k$.
An interpretation is that we use $\ln q$ nats to represent the branches of node $k$ due to the restriction of the alphabet size, \ie, there are at most $q$ branches in a $q$-ary source code.
However, minimally it can be done by $h_k$ nats.
So, $\ln q-h_k$ is the amount of nats wasted at node $k$.

In the decoding tree of a multi-channel tree-decodable code, an internal node corresponds to a channel.
Let $\alpha_k$ be the alphabet size of the channel node $k$ corresponds to.
The number of branches of node $k$ is $\alpha_k$.
We use $\ln \alpha_k$ nats to represent the $\alpha_k$ branches at node $k$ but minimally it can be done by $h_k$ nats.
Therefore, we have the following straightforward extension of local redundancy.

\begin{definition}
	The local redundancy $r_k$ of an internal node $k$ is $r_k := s_k(\ln \alpha_k - h_k)$ nats.
\end{definition}

Denote by $\mathcal{I}$ the index set of the internal nodes. 
Let $L$ be the expected codeword length of a multi-channel tree-decodable code measured in nats.
We can show that $H(Z) = \sum_{k \in \mathcal{I}} s_k h_k$ by a conditional entropy argument.
On the other hand, we can show that $L = \sum_{k \in \mathcal{I}} s_k \ln \alpha_k$ by a weighted bookkeeping argument.
Then, by considering the difference $L - H(Z)$, we arrive at a multi-channel generalization of the local redundancy theorem.

\begin{lemma} \label{lem:red_entropy}
	$H(Z) = \sum_{k \in \mathcal{I}} s_k h_k$.
\end{lemma}

\begin{IEEEproof}
	Let $d$ be the height of the decoding tree.
	An outcome of $Z$ can be represented by a path from the root node to a leaf node of the decoding tree.
	Let $S_i$ be the random variable for the node having depth $i$ which leads to the outcome of $Z$.
	If the outcome of $Z$ is reached, then the random variables for the larger depths are deterministic, \ie, they have zero entropy.
	Then, $Z = (S_0, S_1, \ldots S_d)$ and $S_0 \to S_1 \to \ldots \to S_d$ forms a Markov chain.
	As $S_0$ must be the root node, we have $H(S_0) = 0$.
	Let $\mathcal{I}_i$ be the index set of the internal nodes having depth $i$.
	We have $H(S_i | S_{i-1}) = \sum_{k \in \mathcal{I}_{i-1}} \Pr(S_{i-1} = k) H(S_i | S_{i-1} = k) = \sum_{k \in \mathcal{I}_{i-1}} s_k h_k$.
	Hence, $H(Z) = \sum_{i = 0}^d H(S_i | S_{i-1}) = \sum_{k \in \mathcal{I}} s_k h_k$.
\end{IEEEproof}

\begin{lemma} \label{lem:red_length}
	$L = \sum_{k \in \mathcal{I}} s_k \ln \alpha_k$.
\end{lemma}

\begin{IEEEproof}
	Each internal node $k$ uses $\ln \alpha_k$ nats to represent the $\alpha_k$ branches, which means that the node adds $\ln \alpha_k$ nats to the codeword length of each leaf.
	That is, the node increases the expected codeword length by $s_k \ln \alpha_k$ nats.
	The proof is done by summing up all the increments made by the internal nodes.
\end{IEEEproof}

\begin{theorem}[Local Redundancy Theorem]
	The redundancy of a multi-channel tree-decodable code is $L - H(Z) = \sum_{k \in \mathcal{I}} r_k$ nats.
\end{theorem}

\begin{IEEEproof}
	By \cref{lem:red_entropy,lem:red_length}, we have $L-H(Z) = \sum_{k \in \mathcal{I}} s_k (\ln \alpha_k - h_k) = \sum_{k \in \mathcal{I}} r_k$.
\end{IEEEproof}

\section{Multi-Channel Huffman Procedure}
\label{sec:huffman}

Let $\{p_1, p_2, \ldots, p_m\}$ be the probability distribution of the source random variable $Z$.
Without loss of generality, assume that $p_1 \leq p_2 \leq \ldots \leq p_m$.
It is well-known that the Huffman procedure \cite{huffman} allows us to efficiently construct an optimal $q$-ary single-channel source code, called a $q$-ary Huffman code, for the information source $Z$ \cite{linear_huffman}.
Every iteration of the procedure selects the smallest $q$ probability masses and merges them into one mass, \ie, the smallest $q$ probability masses are removed from the multiset and then their sum is added back to the multiset.
The codeword lengths for the selected masses are increased by $\ln q$ nat.
Every merge reduces the size of the multiset by $q-1$.
In order to merge all the probability masses into one single mass, we need to inject $0 \le w < q-1$ \emph{dummy symbols} to the information source, \ie, inject $w$ \emph{dummy masses} with zero probability into the multiset, before we start the procedure.
The value of $w$ is the smallest non-negative integer such that $m+w \equiv 1 \pmod{q-1}$, \ie,
$$w = [q-1-((m-1)\bmod(q-1))]\bmod(q-1).$$
We call the leaves in the decoding tree which are assigned to dummy symbols the \emph{dummy leaves}.

The core idea of the Huffman procedure is that the smaller the probability mass the longer the codeword length, so we can assign codewords of shorter lengths to those source symbols which are more likely to appear.
This idea is also valid in the multi-channel case.
A traditional proof, e.g., \cite[Lem.~4.15]{info_book}, can be used to prove the following lemma.

\begin{lemma} \label{lem:shorter}
	In an optimal multi-channel uniquely decodable code, codewords with shorter lengths are assigned to larger probabilities.
\end{lemma}

\begin{IEEEproof}
	Suppose in an optimal code we have codeword lengths $\ell^a > \ell^b$ for the probabilities $p_a > p_b$ respectively. 
	By exchanging the codewords assigned to these two probabilities, the expected codeword length of the code is changed by $(p_a \ell^b + p_b \ell^a)-(p_a \ell^a + p_b \ell^b) < 0$, which contradicts to the optimality of the code.
\end{IEEEproof}

When we adopt the Huffman procedure to a multi-channel source code, we have to decide how many probability masses we should merge with the constraint that the number of probability masses to be selected is $q_i$ for some $i \in [n]$.
Note that the procedure can only produce tree-decodable codes.

\begin{definition}
	A \emph{merge sequence} is a finite sequence where the $i$th term is the number of probability masses (including dummy masses) to be merged by the Huffman procedure in the $i$th iteration.
\end{definition}

When we want to merge a certain number of masses in an iteration but there is more than one channel that can be used,
\ie, they have the same alphabet size,
we can arbitrarily choose a channel as the expected length is not affected and the code is still a prefix code. 

The merge sequences for \cref{fig:optimal_eg1,fig:optimal_eg2} are $(2, 3)$ and $(3,2)$ respectively.
We can see that it is not necessary to merge a smaller or larger number of masses in an iteration.
The wrong decision may produce a code which is not optimal.

If a merge sequence $(m_1, m_2, \ldots)$ is given, then the number of dummy symbols $w$ can be implicitly derived by $w = \sum_i (m_i-1)+1-m$.
When the merge sequence is undetermined, however, it is unclear that how many dummy symbols are needed exactly.
Nevertheless, we are able to derive an upper bound of the number of required dummy symbols.
For some specific values of $(q_1,\ldots,q_n)$, e.g., $(q_1,q_2) = (2,3)$, the upper bound forces $w$ to be $0$.

We now state two intermediate lemmas regarding the dummy leaves in an optimal decoding tree, followed by the theorem about the upper bound on the number of required dummy symbols.

\begin{lemma} \label{lem:dummy1}
    There exists an optimal decoding tree where the siblings of a dummy leaf are also leaves.
\end{lemma}

\begin{IEEEproof}
    Consider an optimal decoding tree. If the siblings of a dummy leaf are also leaves, then we are done.
    Otherwise, suppose there exists a dummy leaf such that one of its siblings is an internal node.
    Suppose there exists a non-dummy leaf in the sub-tree under this internal node. Then swapping the dummy leaf with this non-dummy leaf reduces the expected length of the corresponding code. Since we start with an optimal tree, this cannot happen.
    We can therefore assume that the leaves of the sub-tree under this internal node are all dummy leaves, then replacing the internal node with a dummy leaf also gives an optimal decoding tree.
    By repeating this procedure, we obtain an optimal decoding tree where the siblings of a dummy leaf are also leaves.
\end{IEEEproof}

\begin{lemma} \label{lem:dummy2}
	There exists an optimal decoding tree where all the dummy leaves are siblings.
\end{lemma}

\begin{IEEEproof}
	Consider an optimal decoding tree.
    Suppose the set of all dummy leaves are siblings, then we are done.
    Otherwise, suppose there exists two dummy leaves which are not siblings.
    By \cref{lem:dummy1}, the siblings of them are all leaves respectively.

    Consider one of these two dummy leaves. If all siblings of this dummy leaf are also dummy leaves, then we can construct another optimal decoding tree by replacing the parent of the dummy leaf by a dummy leaf.
    Therefore, without loss of generality, we can assume that some siblings of both dummy leaves are not dummy leaves.

    We next argue that these two sets of siblings all have the same generalized depth. Suppose not, then swapping a non-dummy leaf of higher generalized depth with a dummy leaf of lower generalized depth reduces the expected length of the corresponding code. Since we start with an optimal tree, this cannot happen.
    We can therefore assume that the two sets of siblings all have the same generalized depth.

    Since the two sets of siblings all have the same generalized depth, swapping leaves from the two sets do not affect the expected length of the corresponding code. We can therefore swap the leaves in such a way that we pack as many dummy leaves into one set as possible.
    There are two possible outcomes.
    Either one set is rid of dummy leaves, or one set is full of dummy leaves.
    For the latter case, we can use the argument above and replace the parent of these dummy leaves with a dummy leaf, and repeat the subsequent arguments again.
    For the former case, we can repeat the above argument until the set of all dummy leaves are siblings.
\end{IEEEproof}

\begin{lemma}
    \label{lem:dummy3}
    Let $k \in \{2,3,\ldots,q_n\} \cap [m]$ be the number of non-dummy probability masses to be merged in the first round.
    Then the parent node of the $k$ leaves associated to the $k$ merged masses is assigned to class $i^*$, where $q_{i^*}$ is the smallest among $\{q_1,\ldots,q_n\}$ such that $q_{i^*} \geq k$.
    Furthermore, the number of dummy leaves required is $w = q_{i^* } - k$.
\end{lemma}

\begin{IEEEproof}
    By \cref{lem:dummy1,lem:dummy2}, we know that there exists an optimal decoding tree where all dummy leaves are siblings, and all other siblings are non-dummy leaves.
    Let $w$ be the number of dummy leaves, and $k$ be the number of non-dummy leaves in this set of siblings.
    We argue that the parent of these siblings is assigned to class $i^*$, where $q_{i^*}$ is the smallest among $\{q_1,\ldots,q_n\}$ with $q_{i^*} \geq k$, and $w = q_{i^*} - k$.
    Suppose not, then the parent is assigned to some class $i$ with $q_i > q_{i^*}$, and $w = q_i - k > q_{i^*} - k$.
    By changing the class of the parent node from $i$ to $i^*$, and reducing the number of dummy leaves from $q_i - k$ to $q_{i^*} - k$, we obtain a decoding tree whose corresponding code has a lower expected codeword length.
    As we start with an optimal decoding tree, this is impossible.
    We therefore conclude that $w = q_{i^*} - k$.
\end{IEEEproof}

\begin{theorem}
    \label{thm:dummy_needed}
    The number $w$ of dummy symbols needed is bounded by $0 \leq w < \max_{i \in [n]}\{q_i - q_{i-1}\}$ where $q_0 := 1$.
\end{theorem}

\begin{IEEEproof}
    Using the notation in \cref{lem:dummy3}, suppose $i^* > 1$. Note that $k > q_{i^* - 1}$, for otherwise $q_{i^* - 1}$ is smaller than $q_{i^*}$ and yet $k \leq q_{i^*-1}$, violating the definition of $i^*$.
    Suppose otherwise that $i^* = 1$. Note that $k > 1$ for otherwise we can replace the parent of the $k = 1$ node to be ``merged'' by the node itself.
    To summarize, we have $w = q_{i^*} - k < q_{i^*} - q_{i^* - 1} \leq \max_{i \in [n]}\{q_i - q_{i-1}\}$, where $q_0 := 1$.
\end{IEEEproof}

We now know the range of the number of dummy symbols we need to add but not the exact number.
On the other hand, we need to know how to merge these dummy symbols, say, should we merge all the dummy symbols in a single iteration, or should we merge certain number of dummy symbols for a specific channel?
The following lemma describes the structure of a specific optimal tree-decodable code.
Note that there may be other optimal tree-decodable codes which do not meet this structure.
However from the lemma, we know that we can merge all the dummy masses in the first iteration.

\begin{lemma} \label{lem:last}
	There exists optimal decoding tree where all dummy leaves and codewords assigned to a certain number ($k \in \{2,3,\ldots,q_n\} \cap [m]\}$) of the smallest non-dummy probabilities are siblings. 
\end{lemma}

\begin{IEEEproof}
	Now we continue with the optimal decoding tree stated in \cref{lem:dummy2}.
	Let $i^*$ be the class the parent node of the codeword for $p_1$ belongs to.
	This codeword must have at least one sibling assigned to an non-zero probability or otherwise we can remove the last symbol of the codeword in the $i^*$th channel to reduce the expected codeword length.
	Let $k-1$ be the number of siblings of this codeword which are assigned to non-zero probabilities, where $2 \le k \le q_{i^*}$.
	Let one of these siblings be assigned to $p_j$ where $p_1 \leq p_2 \leq p_j$.
	By \cref{lem:shorter}, we have $\ell^1 \ge \ell^2 \ge \ell^j = \ell^1$, \ie, $\ell^1 = \ell^2 = \ell^j$.
	If $p_j \neq p_2$, then we can swap their assigned codewords so that the codewords for $p_1$ and $p_2$ are siblings without changing the expected codeword length.
	We can repeat the above arguments to show that the $k$ smallest non-zero probabilities are siblings.

	The final step is to show that when there is a dummy leaf, then it is a sibling of the codeword of $p_1$.
	It can be proved by a similar argument that if the dummy leaves are not the siblings of the codeword of $p_1$, then we can swap a dummy leaf with a non-dummy sibling of the codeword of $p_1$ so that the expected codeword length is either unchanged or reduced.
\end{IEEEproof}

At last, we can use the above lemma to argue with induction that one of the merge sequences produces an optimal decoding tree.
That is, we know that the generalized Huffman procedure can indeed produce an optimal decoding tree, but we do not know the exact merge sequence.

\begin{theorem}
	 There exists a merge sequence which can produce an optimal multi-channel tree-decodable code.
\end{theorem}

\begin{IEEEproof}
    Fix $k \in \{2,3,\ldots,q_n\} \cap [m]$.
    By \cref{lem:dummy3}, we can determine $i^* \in [n]$, where $q_{i^*}$ is the smallest among $\{q_1,\ldots,q_n\}$ with $q_{i^*} \geq k$, and $w = q_{i^*} - k$.
    When we merge the $k$ smallest probability masses, we obtain a reduced multiset of probabilities
    $\{\sum_{j=1}^k p_j, p_{k+1}, \ldots, p_m\}$.
	Let $L'_k$ be the expected codeword length of the optimal tree-decodable code for the reduced multiset of probabilities.
	Based on this reduced tree, we can reverse the merge to obtain a tree for the original multiset of probabilities where its expected codeword length $L_k$ is $L_k = L'_k+\sum_{j = 1}^{k} p_j \ln q_{i^*}$.
    Note that since $L'_k$ is the expected codeword length of the optimal tree-decodable code for the reduced multiset of probabilities, $L_k$ is optimal for the original multiset of probabilities, for this specific $k$.
    Since the method of creating the trees conforms to the format in \cref{lem:last}, there must exist one $k$ which gives the overall optimal tree.
	By considering all possible $k$, the expected codeword length of the optimal tree-decodable code for the original multiset of probabilities is $\min_{k = 2}^{\min\{q_n,m\}} L_k$.
	We can apply the above arguments inductively (but fixing $k \in \{q_1,\ldots,q_n\} \cap [m']$ in subsequent rounds where $m'$ is the number of remaining masses) to all the possible reduced multisets of probabilities to conclude that there exists a merge sequence which can produce an optimal multi-channel tree-decodable code.
\end{IEEEproof}

The importance of the above result is that we do not have to try arbitrary merges on the probabilities: we only combine the smallest probabilities in each merge.
This way, the search space is greatly reduced.

\section{Approximate Multi-Channel Huffman Procedure}
\label{sec:approx}

It is not known whether an optimal tree-decodable code is an optimal prefix code except for the $2$-channel case.
If we further restrict ourselves to optimal tree-decodable codes, the multi-channel Huffman procedure can be applied.\footnote{A testing tree like the one discussed in \cite{noiseless} requires a source symbol to be the specific branches of the internal nodes. Also, the cost of an internal node is independent of the number of branches. That is, constructing an optimal tree-decodable code is different from constructing an optimal testing tree.}
However, the procedure raises another issue that in each iteration we have to decide how many masses are merged.
Trying all possible merge sequences produces 
exponential number of
decoding trees which is inefficient.
Take $(2,3)$-ary codes as an example, the number of possible merge sequences for $m$ probabilities is the $m$th Fibonacci number. 

One way to reduce complexity is to prune the merge sequences according to some metric.
Specifically, we start by trying all possible choices in each iteration and keeping track of the resulting reduced multiset sizes.
Whenever we obtain two merge subsequences producing reduced multisets of the same size, we eliminate the under-performing one according to some metric and continue with the remaining one.
This strategy avoids unfair comparison between multisets of different sizes. For example, we should compare the reduced multisets generated by the merge sequences $(2,2)$ and $(3)$ as both of them reduced the number of masses by $2$.
In the following we show that using this strategy with several natural metrics is sub-optimal.

Given the probabilities $\{0.13,$ $0.199,$ $0.212,$ $0.217,$ $0.242\}$, we apply the Huffman procedure to generate a $(2,3)$-ary tree-decodable code.
Note that we are worse off choosing the $2$nd channel in the 1st iteration and merging a dummy mass with $2$ non-dummy masses, since we can instead choose the $1$st channel to merge the $2$ non-dummy masses which results in a shorter expected codeword length.
Therefore in the example here, we do not need to consider dummy masses.
The candidate optimal merge sequences are $(2,2,2,2)$, $(2,2,3)$, $(2,3,2)$, $(3,2,2)$, and $(3,3)$, while the winner is $(3,2,2)$.

\begin{table}
	\centering
	\caption{Prune by Redundancy}
	\label{tab:red}
	\setlength\tabcolsep{3.5pt}
	\begin{tabular}{r|c|c|c|c}
		merge & \multicolumn{4}{c}{number of remaining masses}\\
		sequence & $4$ & $3$ & $2$ & $1$\\ \midrule
		(2,2,2,2) & \textbf{0.0072895611} & \textbf{0.0073186993} & \cellcolor[gray]{0.8}0.0139724304 & \cellcolor[gray]{0.8}0.024088589\\
		(2,2,3) & \textbf{0.0072895611} & \textbf{0.0073186993} &  & 0.0337666569\\
		(2,3,2) & \textbf{0.0072895611} &  & \textbf{0.0084312615} & \cellcolor[gray]{0.8}0.0681102540\\
		(3,2,2) & & \cellcolor[gray]{0.8}0.0113528472 & \cellcolor[gray]{0.8}0.0120340121 & \cellcolor[gray]{0.8}\textbf{0.0153997900}\\
		(3,3) & & \cellcolor[gray]{0.8}0.0113528472 & \cellcolor[gray]{0.8} & \cellcolor[gray]{0.8}0.1027103422\\
	\end{tabular}
\end{table}

\begin{table}
	\centering
	\caption{Prune by Expected Codeword Description Length}
	\label{tab:exp}
	\setlength\tabcolsep{3.5pt}
	\begin{tabular}{r|c|c|c|c}
		merge & \multicolumn{4}{c}{number of remaining masses}\\
		sequence & $4$ & $3$ & $2$ & $1$\\ \midrule
		(2,2,2,2) & \textbf{0.2280454224} & \textbf{0.5254055629} & 0.9211926030 & 1.6143397835\\
		(2,2,3) & \textbf{0.2280454224} & \textbf{0.5254055629} &  & \cellcolor[gray]{0.8}1.6240178515\\
		(2,3,2) & \textbf{0.2280454224} &  & \cellcolor[gray]{0.8}0.9652142681 & \cellcolor[gray]{0.8}1.6583614487\\
		(3,2,2) & & \cellcolor[gray]{0.8}0.5943492482 & \cellcolor[gray]{0.8}\textbf{0.9125038040} & \cellcolor[gray]{0.8}\textbf{1.6056509846}\\
		(3,3) & & \cellcolor[gray]{0.8}0.5943492482 & \cellcolor[gray]{0.8} & \cellcolor[gray]{0.8}1.6929615368\\
	\end{tabular}
\end{table}

\begin{table}
	\centering
	\caption{Prune by Entropy}
	\label{tab:ent}
	\setlength\tabcolsep{3.5pt}
	\begin{tabular}{r|c|c|c|c}
		merge & \multicolumn{4}{c}{number of remaining masses}\\
		sequence & $4$ & $3$ & $2$ & $1$\\ \midrule
		(2,2,2,2) & \textbf{1.3694953333} & \cellcolor[gray]{0.8}1.0721643310 & \cellcolor[gray]{0.8}0.6830310220 & \cellcolor[gray]{0.8}\textbf{0.0000000000}\\
		(2,2,3) & \textbf{1.3694953333} & \cellcolor[gray]{0.8}1.0721643310 & \cellcolor[gray]{0.8} & \cellcolor[gray]{0.8}\textbf{0.0000000000}\\
		(2,3,2) & \textbf{1.3694953333} &  & \textbf{0.6334681881} & \textbf{0.0000000000}\\
		(3,2,2) & & \textbf{1.0072547937} & \cellcolor[gray]{0.8}0.6897814027 & \cellcolor[gray]{0.8}\textbf{0.0000000000}\\
		(3,3) & & \textbf{1.0072547937} &  & \textbf{0.0000000000}\\
	\end{tabular}
\end{table}

\begin{table}
	\centering
	\caption{Prune by Expected Resultant Codeword Length $+$ Entropy}
	\label{tab:exp+ent}
	\setlength\tabcolsep{3.5pt}
	\begin{tabular}{r|c|c|c|c}
		merge & \multicolumn{4}{c}{number of remaining masses}\\
		sequence & $4$ & $3$ & $2$ & $1$\\ \midrule
		(2,2,2,2) & \textbf{1.5975407557} & \textbf{1.5975698939} & \cellcolor[gray]{0.8}1.6042236250 & \cellcolor[gray]{0.8}1.6143397835\\
		(2,2,3) & \textbf{1.5975407557} & \textbf{1.5975698939} &  & 1.6240178515\\
		(2,3,2) & \textbf{1.5975407557} &  & \textbf{1.5986824562} & \cellcolor[gray]{0.8}1.6583614487\\
		(3,2,2) & & \cellcolor[gray]{0.8}1.6016040418 & \cellcolor[gray]{0.8}1.6022852068 & \cellcolor[gray]{0.8}\textbf{1.6056509846}\\
		(3,3) & & \cellcolor[gray]{0.8}1.6016040418 & \cellcolor[gray]{0.8} & \cellcolor[gray]{0.8}1.6929615368\\
	\end{tabular}
\end{table}

During the Huffman procedure, we record the redundancy (sum of local redundancies) and the expected codeword length of the already constructed subtree(s) in \cref{tab:red,tab:exp}.
We also record the entropy of the reduced multisets of probabilities in \cref{tab:ent}.
\cref{tab:exp+ent} shows the entry-wise sum of \cref{tab:exp,tab:ent}.
The values in \cref{tab:ent,tab:exp+ent} are the (lower and upper) bounds of the expected codeword length of the not-yet-constructed part and the expected codeword length of the resultant code respectively (by \cref{thm:entropy,thm:source}).
All the values are measured in nats.

The numbers in bold are the minimums of the corresponding columns.
For each table, we compare the steps of different merge sequences when the numbers of remaining masses are the same.
If a merge sequence does not attend the minimum during a comparison, we prune the merge sequence away and highlight the corresponding cells in gray.
The merge sequence with a white cell in the rightmost column is the output of the procedure for the specific metric of the table.
We can see that none of these tables produce the optimal $(3,2,2)$.

We now propose a straightforward suboptimal code construction which can guarantee a redundancy no larger than the one of a single-channel Huffman code.
We follow a similar idea used in \cref{tab:exp+ent} except that, instead of adding the entropy of the reduced multiset of probabilities to the expected codeword length of the already constructed subtree(s), we add the smallest expected codeword length of the single-channel Huffman codes constructed for the reduced multiset of probabilities on different channels.
This sum is the expected codeword length of a code which can be constructed explicitly.
Note that a single-channel Huffman code falls in one of the merge sequences, so that we can ensure the code we produce this way has a redundancy no larger than a single-channel Huffman code.

\begin{table}
	\centering
	\caption{Suboptimal Code Construction}
	\label{tab:sub}
	\setlength\tabcolsep{3.5pt}
	\begin{tabular}{r|c|c|c|c}
		merge & \multicolumn{4}{c}{number of remaining masses}\\
		sequence & $4$ & $3$ & $2$ & $1$\\ \midrule
		(2,2,2,2) & \textbf{1.6143397835} & \cellcolor[gray]{0.8}1.6143397835 & \cellcolor[gray]{0.8}1.6143397835 & \cellcolor[gray]{0.8}1.6143397835\\
		(2,2,3) & \textbf{1.6143397835} & \cellcolor[gray]{0.8}1.6143397835 & \cellcolor[gray]{0.8} & \cellcolor[gray]{0.8}1.6240178515\\
		(2,3,2) & \textbf{1.6143397835} &  & \cellcolor[gray]{0.8}1.6583614487 & \cellcolor[gray]{0.8}1.6583614487\\
		(3,2,2) & & \textbf{1.6056509846} & \textbf{1.6056509846} & \textbf{1.6056509846}\\
		(3,3) & & \textbf{1.6056509846} &  & \cellcolor[gray]{0.8}1.6929615368\\
	\end{tabular}
\end{table}

\section{Conclusion and Future Directions}
\label{sec:conclude}

In this paper, we showed that it is possible for a multi-channel source code to achieve a better compression than an optimal single-channel source code.
We also presented the multi-channel analog of some classical results in single-channel source coding.
Some future research directions on multi-channel source coding are:
\begin{enumerate}
	\item Is an optimal multi-channel tree-decodable code an optimal prefix (or uniquely decodable) code?
	\item Is there a metric with which the pruning strategy results in an optimal multi-channel Huffman code?
	\item How to extend adaptive, canonical and run-length Huffman codes into their multi-channel version?
	\item How to restrict the expected codeword length for each channel and to balance the usage of each channel?
\end{enumerate}

\bibliographystyle{IEEEtran}
\bibliography{prefix-bib}

\end{document}